\documentclass[twoside,twocolumn,10pt]{article}

\usepackage{wscg}           
\RequirePackage{ifpdf}
\ifpdf
 \RequirePackage[pdftex]{graphicx}
 \RequirePackage[pdftex]{color}
\else
 \RequirePackage[dvips,draft]{graphicx}
 \RequirePackage[dvips]{color}
\fi

\usepackage{nopageno}       

\title{A User-centered Approach for Optimizing Information Visualizations}

\author{
\parbox{0.75\textwidth}{\centering
David Baum, Pascal Kovacs, Ulrich Eisenecker and Richard M\"uller\\[1mm]
Leipzig University\\
Grimmaische Strasse 12\\
04109 Leipzig, Germany\\[1mm]
[baum, kovacs, eisenecker, rmueller]@wifa.uni-leipzig.de
}
}

\usepackage{url}
\makeatletter
\def\Uslash{\mathbin{\mathchar`\/}\@ifnextchar{/}{\kern-.15em}{}}
\g@addto@macro\UrlSpecials{\do \/ {\Uslash}}
\def\Ucolon{\mathbin{\mathchar`:}\@ifnextchar{/}{\kern-.1em}{}}
\g@addto@macro\UrlSpecials{\do : {\Ucolon}}
\makeatother

\begin{document}
    
\twocolumn[{\csname @twocolumnfalse\endcsname
\maketitle  


\begin{abstract}
    The optimization of information visualizations is time consuming and expensive. 
    To reduce this we propose an improvement of existing optimization approaches based on user-centered design, focusing on readability, comprehensibility, and user satisfaction as optimization goals.
    The changes comprise
    (1) a separate optimization of user interface and representation, 
    (2) a fully automated evaluation of the representation, and
    (3) qualitative user studies for simultaneously creating and evaluating interface variants.
    On the basis of these results we are able to find a local optimum of an information visualization in an efficient way.
    


\end{abstract}        

\subsection*{Keywords}
Evaluation, Information Visualization, Optimization, Usability, User-centered design

\vspace*{1.0\baselineskip}
}]

\section{Introduction}
\label{sec:introduction}

\copyrightspace


%

Over the last years, a considerable number of visualizations has been presented \cite{Caserta2011, Ltifi2009, Liu2014, Teyseyre2008, Landesberger2011, Zudilova-Seinstra2014}.
The benefit of a specific visualization depends on many factors, such as addressed stakeholder (e.g. project manager, analyst, scientist, or developer), the chosen methods of representation and interaction, and the supported tasks \cite{Liu2014, Landesberger2011}.
Because of the number of factors and their connections evaluating visualizations is a big challenge.
Nevertheless, in most cases more time is spent on developing entirely new visualizations than to evaluate them and some of them have not been evaluated at all \cite{Wettel2011,Teyseyre2008}. 

Empirical quantitative studies are an established type of evaluation and can prove that one visualization is superior over another one.
However, planning, conducting and analyzing such a quantitative study is difficult, time-consuming and causes a huge effort \cite{Andrews2006, Chen2000, Kerren2008, Lam2014, Plaisant2004}.
Especially, recruiting a sufficient number of participants is hard if they have to meet certain criteria such as specific profession (e.g. software developer with industrial experience).
Tasks are another critical aspect in such studies, because simple tasks are easier to create but are not representing a real world scenario, which is a thread to the external validity\cite{Landesberger2011}. 
Complex tasks on the other side are more difficult to create, because of the higher risk of misinterpretation by the participants.
Despite of these difficulties, a quantitative study may lead to significant results, but often gives not enough insight into the details, why a visualization is superior \cite{Lam2014}.
Furthermore, the choice which visualizations or visualization variants should be investigated is a critical part of what the results are useful for, but in most cases it is not exactly reasoned. 

These obstacles apply to evaluation of visualizations in general and even more to their optimization, because to achieve a satisfying visualization several improvements and therefore evaluations have to be done.
Thus, not only a single visualization has to be evaluated but also several variants differing in representation details and interaction options.
Due to the complexity of most visualizations the amount of possible variants is far too high for evaluating every variant. 
Therefore, it is necessary to apply reasonable strategies to reduce the number of variants to be evaluated down to a manageable number.

In this paper we present our approach for the optimization of visualizations regarding readability, comprehensibility, and user satisfaction, derived on our experience of evaluating software visualizations. 
We combine computational, qualitative and quantitative methods into a well-structured and repeatable process, based on existing processes for user-centered design (UCD) and considering the specific characteristics of information visualizations. 
By adopting this process a researcher can reduce time and effort finding a local optimum in an efficient heuristic way to improve any visualization.

%
%
%
%
%
%
%
%

\section{Approach}
\label{sec:approach}

\begin{figure*}[htb]
    \centering
    \includegraphics[width=\textwidth]{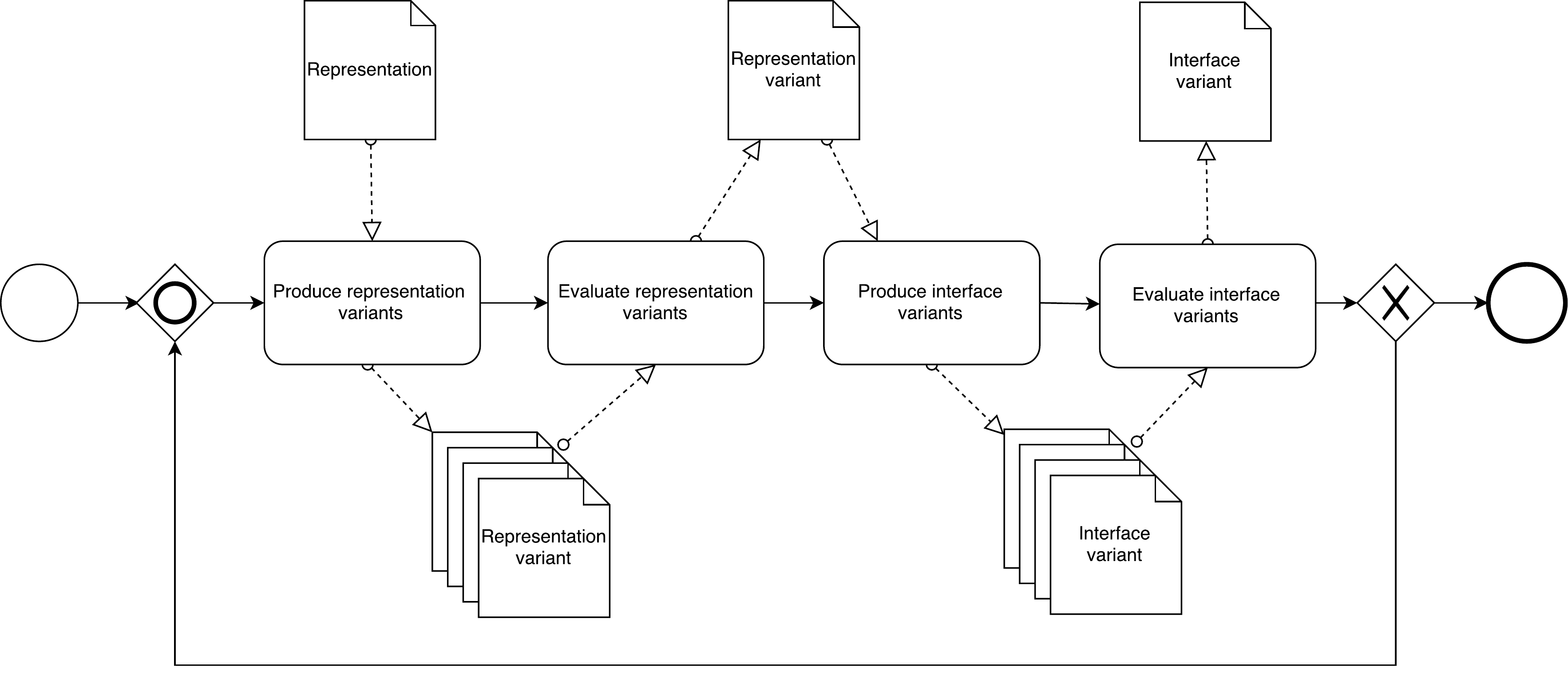}
    \caption{Optimization process for information visualizations}
    \label{fig:process}
\end{figure*}

User-centered approaches are usually based on at least four iterative steps \cite{Poppe2006}:
\begin{enumerate}
 \item identify need for UCD
 \item produce design solutions
 \item evaluate designs against requirements
 \item final design
\end{enumerate}


Compared to existing approaches we alter steps 2 and 3 by optimizing user interface (UI) and representation separately.
Thereby, for each part of the optimization the most efficient and suitable method can be chosen.
We propose an aesthetics-based approach for representation optimization (section \ref{sec:aesthetics}) and user studies for UI optimization (section \ref{sec:expertinterviews}) as shown in figure~\ref{fig:process}.
Like existing UCD approaches, the whole process is repeated until a certain criterion is reached \cite{Wassink2009}.
Depending on the motivation of the optimization this can be, e.g., a targeted deadline or the detection of merely non-significant improvements.



Munzner \cite{Munzner2009a} introduced a four-layered model for visualization design and validation.
According to this model, our approach refers to the layer \emph{encoding/interaction technique design}.

\section{Optimize Representation}
\label{sec:aesthetics}

Aesthetics are visual properties of a representation that are observable for human readers as well as measurable in an automated way \cite{Baum2015}.
Some of them affect the readability and comprehensibility significantly, either in a positive or in a negative way \cite{Purchase1997}.
The effects on user performance can be measured in a quantitative study using the time a user needs to solve a task and the number of errors he makes \cite{Huang2014}.
Based on aesthetics, representations that are optimized for readability and comprehensibility can be designed.

As aesthetics emerge from the properties of the depicted elements, such as color, shape, size, and positioning, they are specific for every representation \cite{Bennett2007, Purchase2002}.
For some basic representations like node-link diagrams aesthetics and their influence on readability and comprehensibility are well-understood \cite{Bennett2007}.
In this case the process of optimization becomes easier, since some part of the work is already done.
The gathered knowledge about aesthetics can be reused in further iterations.
Hence, the effort is reduced with every iteration and quantitative studies might even become obsolete.

\subsection{Produce Representation Variants}

The previous work of Baum \cite{Baum2015} describes how the repertory grid technique can be used to identify relevant aesthetics for any representation in a structured and reproducible way.
The resulting list of aesthetics is narrowed down by two requirements that have to be fulfilled.
First, no information may be lost; second, there must be a significant effect on user performance.
A solely aesthetics-based optimization of readability is not meaningful if the changes imply an adulteration of the visualized content.
For example, a layout algorithm might convey information via the order of the depicted elements.
If this order is changed, e.g., to reduce space consumption, the result may be more readable but some information is tampered.
%
Further, it is unlikely that all identified aesthetics have a significant effect on user performance, based on the experiences with node-link diagrams \cite{Ware2002}.
To reveal the relations between aesthetics and user performance quantitative studies are still required.
Every examined visualized data set is based on the same visualization but holds different values for one or multiple aesthetics.
Measuring the time needed by a user to solve a task and the number of errors made while doing so yields two important findings.
First, the aesthetics that have a significant effect on user performance; second, the weighting of those aesthetics since they differ in their impact.

Eventually, one or more variants of the original representation can be created with respect to the most influential aesthetics, e.g., by applying another layout algorithm.
Except during the first iteration the results of the user studies can be used as additional source of information.
Producing variants still requires the creativity of the researcher since aesthetics only determine the goal of the optimization but not how it can be achieved.
For example, our approach does not help to develop completely new layout algorithms, but aesthetics provide assessment criteria for automatic evaluation. 

\subsection{Evaluate Representation Variants}

Aesthetics allow a fully automatized evaluation \cite{Purchase1997}.
For every created variant its effect on readability and comprehensibility can be automatically calculated by making use of the gathered information.
Hence, the evaluation is very efficient and even a large amount of variants can be evaluated without difficulty.
The outcome of the evaluation is a representation variant that will be further optimized.

\section{Optimize Interaction}
\label{sec:expertinterviews}

The interaction between the user and a visualization is realized through the UI, which is a complex combination of interaction techniques (ITEC).
Yi et al. \cite{Yi2007} define ITECs in information visualization as \emph{
"[...] the features that provide users with the ability to directly or indirectly manipulate and interpret representations".}
To categorize ITECs they propose a taxonomy of seven categories: \emph{select, explore, reconfigure, encode, abstract/elaborate, filter, and connect}.
Hence, evaluating the interaction with a UI via ITECs could be done in four different levels of detail, from low to high, by 
\begin{itemize} 
\item comparing full UIs against each other,
\item integrating ITECs in the UI,
\item pairwise comparisons of ITECs of one category, and
\item scrutinizing details of a single ITEC.
\end{itemize}
With the target of optimizing the interaction as a whole, a quantitative evaluation in one of these levels is not suitable, because either the reasons why one UI is superior over another UI can not be identified or the context of the target domain is lost when evaluating only the details of one ITEC.
A quantitative evaluation of all four levels is also not feasible, because of the huge effort and the difficulties, even when comparing only two variants per level\cite{Lam2008}. 
Furthermore, the space of possible variants is huge, thus choosing the variants for further evaluation and improvement is a critical part.

Therefore, we propose iterative qualitative user studies in a within-subject design as a heuristic to find a local optimum in the huge space of possible UI variants.
One iteration consists of a couple of runs, where every participant solves a set of randomized tasks using an optimized representation variant and more than one UI variant.
Each UI variant differs in at least one detail of ITECs, e.g., one variant has  zoom by mouse wheel to the position of the cursor, the other one zooms by double click on an element, and a third one zooms twice as fast as the second one using an addition button.
The first iteration starts with some UI variants chosen by the researcher, which are derived from his own ideas or by other visualizations or guidelines.
Further iterations may contain subsequent UI variants triggered by analyzing the feedback of participants.
Additionally, tasks may be altered, bugs in the visualization can be fixed, and ideas for representational variants could be identified, which will be used during representation optimization.
If the optimization process is terminated a final UI is derived from the evaluation of the investigated UI variants.

To get as much detail about the interaction as possible, qualitative data is collected about each UI variant and also about the tasks and their descriptions.
Therefore, the feedback and questions during and after each task execution as well as the instructions and observations of the experimenter are gathered.
The user actions including their timestamps and the time- and error-rate for the solved task are recorded too. 
However, with respect to the bias of giving feedback during the task, the possible misinterpretation of the task description and the variance in user skills coupled with a low number of participants, the time- and error-rate have to be interpreted with caution.
After solving the full task set, the participant eventually has to rank all UI variants from best to worst. 
The ranking of all participants of one iteration shows which UI variants support the set of tasks better than others. 
Furthermore, it may give hint to factors explaining the improvements.

Beside changing UI variants, tasks and their descriptions can be changed or improved between iterations as well, because designing tasks is not straightforward. 
Too simple tasks, e.g., \emph{identify the largest element}, are not suitable as a real world task for visualization analysis. 
On the other hand, a complex task is more difficult to explain, may be misinterpreted by the participant, or needs too much time to be solved \cite{North2006}.
Thus, creating and describing a perfect set of complex tasks from scratch is nearly impossible.
To overcome this problem a pilot study is an established way to find weaknesses in tasks and their descriptions.
However, the possible task modifications found this way are only a subset and every modification can lead to new weaknesses.
Hence, an iterative improvement is a better solution to optimize the tasks.
By analyzing the instructions and observations of the experimenter as well as the questions and feedback of the participants the researcher draws several conclusions about the comprehensibility and feasibility.
As a result, the complexity of the tasks can be reduced, the descriptions can be remastered, or entire tasks can be replaced.

\subsection{Produce Interface Variants}

To produce new variants the within-subject design is chosen to encourage the participants to think about the differences between the variants.
Therefore, the participants feedback and questions are collected during the whole process and associated to the following categories:
\begin{itemize} 
\item advantages of variants
\item disadvantages of variants
\item improvements for variants
\item ideas for new variants 
\end{itemize}
By summarizing and interpreting the categorized statements and their rate the researcher draws several conclusions about possible changes.
This interpretation process is not of straightforward structure because the researcher and his or her freedom to design the UI is also part of it.
For example, the number of gathered disadvantages for one variant may lead the researcher to an idea how to improve this variant to overcome this disadvantages.
So the freedom in designing the UI using the qualitative data is the crucial part to find a local optimum in the huge space of possible variants.
Nevertheless, the researcher should pay attention to explicitly record his or her decision with respect to further planning of the optimization process.
The result of this analysis is an overview following possible changes for the next iteration, weighted by the potential contribution to the effectiveness of the UI: 
\begin{itemize} 
\item adding a complete new variant
\item adding an altered existing variant
\item adding a combination of existing variants
\end{itemize}
Attention should be paid to the differences between the variants in one iteration. 
If they differ in every possible detail of the UI or the ITECs the participants may become confused and the comparison of the variants may not lead to relevant feedback.
This would also lead to very long instruction-phases with broad tutorials to explain each variant in detail.
Hence, the changes should at least be focused on one category of ITECs, e.g., explore or connect.
However, the level of detail in the differences should be taken into account too.
The details of the ITECs and their integration into the UI should be investigated after evaluating if and under which conditions a certain ITEC is superior.

Depending on the amount of existing variants and the size of the task set one or more variants can be added for the next iteration.
To consider a bigger amount of variants new tasks could be added too, but with respect to the overall length for solving all tasks of the set.
On the other side, old variants can be removed if they are ranked low by the participants or have many disadvantages.

\subsection{Evaluate Interface Variants}

The evaluation of the variants is mainly driven by the user satisfaction, recorded as the ranking from best to worst for all variants after solving the complete task set.
To get a ranking for the whole iteration the medians for each variant are computed.
An aggregated ranking for all investigated variants in all iterations is built by computing the medians of this iteration rankings, so new variants will not be outnumbered by older ones.
This way less effective UI variants are identified and can be excluded from the next iteration.
If the process of optimization comes to an end a final variant out of the remaining variants has to be derived.
Beside the ranking the circumstances why and when a variant is more effective than another one are also part of this final decision.
Therefore, at least all the best ranked variants are investigated further as final candidates by analyzing the advantages and disadvantages as well as comparing the quantitative data of time- and error-rate or the user actions.
This may lead to the following four cases:
\begin{enumerate}  
\item Interpreting the advantages and disadvantages can lead to the conclusion that a final candidate is only superior for a specific type of task. 
In this case either a new variant should be built upon this insight or, if not possible, all these remaining candidates should be integrated in the final UI with respect to aesthetics of the graphical elements of the UI \cite{Zen2014}.
Thus the user can decide which variant to use for a task. 
\item Computing the relevant statistical parameters of time- and error-rates identifies one final candidate as noticeably superior over the others.
\item One of the candidates has a noticeably lower rate in user actions to solve the tasks than the others.
In a long term usage this candidate should have a higher acceptance by the users.
\item The differences between the candidates are only on a low level of detail, so they could be integrated in the final UI by a configuration option.
\end{enumerate}
If the result of analyzing the final candidates can not be classified as one of these cases either a further investigation by conducting a quantitative study could be done or the researcher eventually has to choose the final UI.

\section{Discussion}
\label{sec:discussion}

In this paper we propose some relevant changes to existing UCD processes to reduce the effort for optimizing visualizations.
Although we were able to apply the process successfully an evaluation against other evaluation approaches is outstanding due to the required effort.
Especially the implementation of the variants is still time-consuming.
Since we consider interaction as a crucial factor of success of a visualization we decided against paper prototyping and similar methods.
To further increase the efficiency, and thereby being able to evaluate more variants, it is essential to at least partially automate the evaluation of UI variants.
However, the current understanding of UI aesthetics is not yet deep enough \cite{Zen2014}.

%
Among others, we use quantitative studies to optimize the representation.
Even though their number is reduced over time, the first iterations might be even more extensive than existing approaches.
However, experience shows that usually many iterations are required and in that case our approach becomes less extensive.

The described approach finds only a local optimum, since it is unfeasible to evaluate all possible variants.
This limitation is common to all optimization processes in the area of information visualization.
However, our approach comes with a highly efficient evaluation.
User studies are used simultaneously for creating and evaluating UI variants in smaller iterations, by analyzing the qualitative data and user ranking.
Then the evaluation of the representation is fully automated.
Thus, we can investigate a much bigger space to find the local optimum.
In turn, the evaluation results are less reliable compared to quantitative studies.
Therefore, we propose to finish the optimization process with a controlled experiment to make sure it was successful.

\section{Related Work}
\label{sec:relatedwork}

Several papers address the methodology of evaluating information visualizations \cite{Carpendale2008, Holmberg2006, Lam2014, Micallef2012, Moere2012, Seriai2014, Tory2005}.
But they only focus on single evaluations, not on an iterative process as described in this paper.
However, iterative optimization is an essential part of UCD. 
Some authors described such user-centered approaches for information visualization \cite{Fernandez2013, Lloyd2011, Wassink2009}.
As we, they try to reduce the resulting effort, e.g., by combining controlled experiments and qualitative methods.
Unfortunately, this is achieved at the expense of a drastically reduced interaction evaluation.
In contrast, we stress the importance of the interaction but still achieve a reduced effort.

\section{Conclusion}
\label{sec:conclusion}

In this paper, we proposed an improved process for optimizing information visualization regarding readability, comprehensibility, and user satisfaction.
Among a heuristic process of finding a local optimum in the huge space of UI variants, we introduced a fully automated evaluation of the representation variants.
Although we were able to apply the process successfully an evaluation against other evaluation approaches is outstanding.

{\small
    \bibliography{david,pascal,references}
}
\bibliographystyle{alpha}

\end{document}